\documentclass[useAMS,usenatbib]{mn2e}

\usepackage{graphicx}

\usepackage{lastpage}

\usepackage{txfonts}

\title[Methanol maser variability in Cepheus A]{6.7\,GHz methanol maser variability in Cepheus A}

\author[M. Szymczak, P. Wolak and A. Bartkiewicz]
{M. Szymczak \thanks{E-mail: msz@astro.uni.torun.pl},
P. Wolak
and A. Bartkiewicz
\\
Centre for Astronomy, Faculty of Physics, Astronomy and Informatics, Nicolaus Copernicus University,\\
 Grudziadzka 5, 87-100 Torun, Poland \\
}

\begin{document}

\date{Accepted 2013 December 19. Received 2013 December 18; in original form 2013 September 6}

\pagerange{\pageref{firstpage}--\pageref{LastPage}} \pubyear{2014}

\maketitle

\label{firstpage}
\begin{abstract}
6.7\,GHz methanol maser emission from the well-studied star-forming region Cepheus\,A 
was monitored with the Torun 32\,m radio telescope. We found synchronized and 
anticorrelated changes of the flux density of the two blueshifted and one redshifted 
maser features for $\sim$30 per cent of 1340\,d of our observations.
Two of those features exhibited high amplitude flux density variations with periods 
of 84$-$87\,d over the last 290\,d interval of the monitoring.  
We also report on two flares of emission at two different redshifted velocities 
completely covered during the whole outburst. These flare events lasted 510$-$670\,d 
and showed a very rapid linear rise and slow exponential decline, which may be caused 
by variability of the seed flux density. The flux density of the two strongest
features dropped by a factor of 2$-$5 on a time-scale $\sim$22 yr, while other features 
have not changed significantly during this period, but showed strong variability 
on time-scales $\la$5 yr.
\end{abstract}


\section{Introduction}
Cepheus A (Cep\,A) is a nearby, 0.70$\pm$0.04\,kpc
\citep{moscadelli09}, massive star-forming region with a bolometric
luminosity of $2.5\times10^4 L_{\sun}$ \citep{evans81}. It contains an
extended ($\sim$1\,arcmin) molecular outflow of complex morphology
that is probably powered by the radio continuum source HW2
\citep{hughes84}. Interferometeric observations of the dust continuum,
free--free emission and several molecular tracers have shown that HW2
is surrounded by circumstellar discs of dust and gas of radii of
$\sim$300 and $\sim$600\,au, respectively (\citealt{patel05};
\citealt{curiel06}; \citealt{jimenez07, jimenez09};
\citealt{torrelles11}). Very long baseline interferometry (VLBI) observations 
of water masers revealed the presence of a slow ($\sim$10--70\,km\,s$^{-1}$) wide-angle
($\sim$102\degr) outflow together with a high-velocity
($\sim$500\,km\,s$^{-1}$) ionized jet with an opening angle of
$\sim$18\degr\, associated with HW2 \citep{torrelles11}. 6.7\,GHz
methanol masers originate in an arc-like structure of size of 1350\,au
(\citealt{sugiyama08a}; \citealt{vlemmings10};
\citealt{torstensson11}). \citet{torstensson11} proposed a model in which the
maser ring has a radial motion of 1.3\,km\,s$^{-1}$, probably infall,
and the methanol maser emission probes the interface between the
accretion flow and the disc. 
Polarimetric observations of the 6.7\,GHz
transition indicated that the magnetic field probably regulates
accretion on to the disc \citep{vlemmings10}.

Single-dish monitoring over a period of 81\,d revealed that the flux
density variability of the blue- and redshifted 6.7\,GHz maser
features in this source is synchronized but anticorrelated
\citep{sugiyama08b}. This finding essentially excludes a hypothesis
that the variability is caused by collisional excitation by a
shock wave from a common powering source. Instead, the negative
correlation can be attributed to  changes in the dust temperature of
the maser regions due to their distances from a common heating source and
its luminosity variations \citep{sugiyama08a}. \citet{galt03} reported
a significant change in the flux density ratio of the two strongest
blueshifted features over two years. The strongest feature in the
spectrum, near $-$2.6\,km\,s$^{-1}$, has dropped from 1410 to 815\,Jy on
a time-scale of 8.1\,yr (\citealt{menten91}; \citealt*{szymczak00}).  In
this paper, we report nearly 4\,yr monitoring observations of Cep\,A at
6.7\,GHz with the aim to expand our understanding of the maser
variability. The results can be used to probe the environment of HW2 
and could further shed light on excitation mechanisms and
physical processes related to the maser emission.

\section{Observations}
Observations of the 6668.519\,MHz methanol line were carried out using
the Torun 32\,m radio telescope.  The target source (RA(J2000) =
22$^{\rm h}$56$^{\rm m}$17\fs9, Dec.(J2000) = 62$^0$01\arcmin49\farcs6)
was monitored from 2009 June to 2013 February at irregular intervals
of 1--7\,d with %( 2009.06.24 to 2013.02.23)
the exception of three gaps of 3--4 weeks in the monitoring campaign
due to other projects of higher priority.  In total, useful data
were obtained at 388 epochs, in the framework of a methanol maser
monitoring programme which is still going on. 
% intervals of 1$-$7 days were most frequent (337/367), only three gaps 3-4 weeks. 388 data points
A dual-channel HEMT receiver was used in a frequency-switching mode. 
The system temperature was about 40\,K  on cold sky.
It was measured every 5\,min by injecting a signal from a noise 
diode of known temperature. The conversion factor of the antenna temperature 
to the flux density for a point source was $\sim$0.14\,K\,Jy$^{-1}$.
The half-power beam width at 6.7\,GHz was 5.5\,arcmin and the rms pointing 
error was 22\,arcsec. The two opposite circular polarizations were measured 
in 4\,MHz bandwidth divided into 4096 channels, resulting in a velocity 
resolution of 0.05\,km\,s$^{-1}$ after Hanning smoothing. 
The typical rms noise level in the spectra after averaging the two 
polarizations was 0.30--0.35\,Jy. The flux density scale 
was derived by continuum observations of source 3C~123, assumed to have 
a flux density of 11.76\,Jy  \citep{ott94}. The 32\,m antenna gain reduction 
at low (15\degr) elevation angle was $\sim$2.5 per cent. 

%%%%%%%%%%%%%%%%%%%%%%%%%%%%%%
\begin{figure}   
\resizebox{\hsize}{!}{\includegraphics[angle=0]{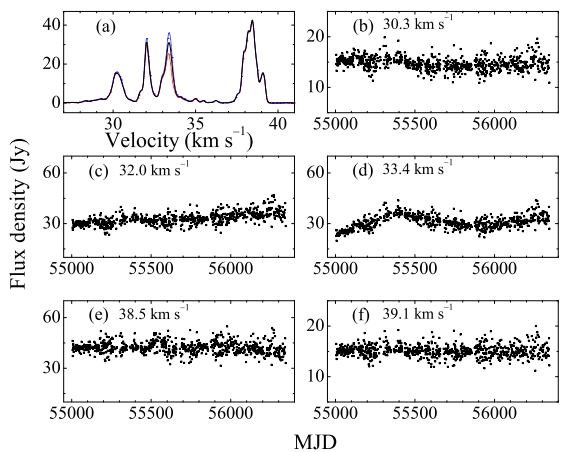}}
\caption{(a) Range of variation in spectrum of G32.745$-$0.076. The solid line 
is the averaged spectrum. The two dotted lines
are the upper and lower envelopes. (b)$-$(f) Time series for strong features.  
\label{g32p7}}
\end{figure}
%%%%%%%%%%%%%%%%%%%%%%%%%%%%%%

The maser source G32.745$-$0.076 previously reported as non-variable 
(\citealt*{caswell95}) within $\sim$5 per cent \citep{szymczak11} 
was regularly observed at 801 epochs in order to check the stability of the system. 
Fig. \ref{g32p7} shows the range of variation in the spectrum of 
G32.745$-$0.076 and the time series of strong features.
The emission features near 30.3, 38.5 and 39.1\,km\,s$^{-1}$ have not 
shown variability in excess of $\sim$8 per cent during the monitoring period. 
The other two features near 32.0 and 33.4\,km\,s$^{-1}$ show high variability 
of about 19 and 37 per cent, respectively.  
Based on the standard deviation of the flux density values of `non-variable' 
features of this source, we estimate that our flux density calibration 
is better than 10 per cent.

To quantify the variability, we used the variability indices $vi_{\rm 1}$ and 
$vi_{\rm 2}$. The first index is given (e.g. \citealt*{aller92}) by

\begin{equation}
 vi_1 = {(S_{\rm max} - \sigma_{\rm max}) - (S_{\rm min} + \sigma_{\rm min})\over (S_{\rm max} - \sigma_{\rm max}) + (S_{\rm min} + \sigma_{\rm min})}
\end{equation}

\noindent
where $S_{\rm max}$ and $S_{\rm min}$ are the highest and lowest
measured flux densities, respectively, and $\sigma_{\rm max}$ and
$\sigma_{\rm min}$ are the uncertainties in these measurements. This
variability index is a measure of the amplitude of the variability of
the source and allows for the effect of measurement uncertainties.
Its value is well determined only when variability is significantly
greater than measurement errors and can be negative for sources with
low signal-to-noise (S/N) ratios \citep{aller92}.

The second variability index is defined (e.g. \citealt{edelson02}) as

\begin{equation}
 vi_2 = {\big(}{1\over N}\sum_{i=1}^N[(S_{\rm i} - \overline{m})^2 - \sigma_{\rm i}^2]{\big)}^{0.5}{\big/}\overline{m} 
\end{equation}
 
\noindent
where $N$ is the total number of observations, $S_{\rm i}$ is 
the observed flux density, $\sigma_{\rm i}$ is the rms level
in the spectrum and ${\overline m}$ is the average flux density. 
This index is a useful measure of variability for spectral channels 
with low S/N ratio. These variability indices are discussed in details 
in \citet{aller92} and  \citet{edelson02}.

\section{Results}
\begin{figure}   
\resizebox{\hsize}{!}{\includegraphics[angle=0]{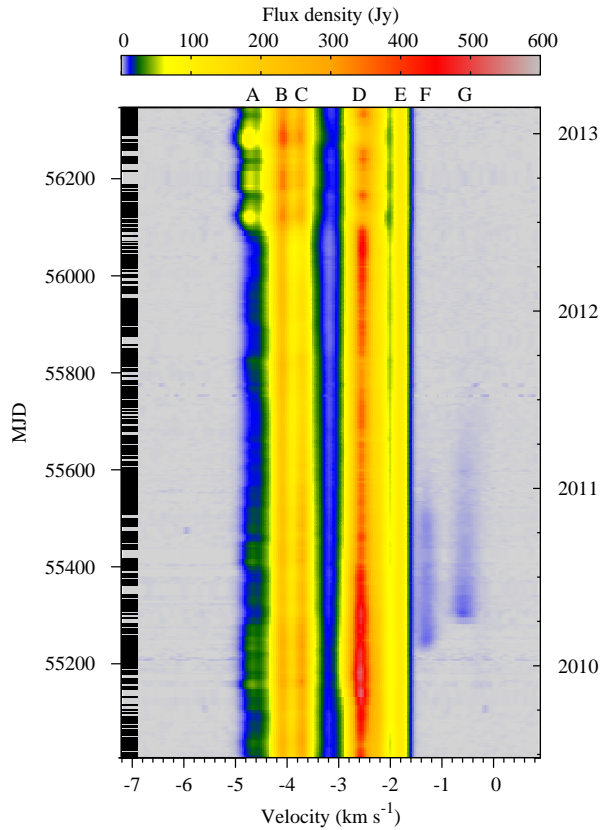}}
\caption{False-colour image of the 6.7\,GHz methanol maser flux 
density over velocity and time for Cep\,A. The velocity scale 
is relative to the local standard of rest. Labels A--G indicate 
the analysed features. The horizontal bars in the left coordinate 
correspond to the dates of the observed spectra.  
\label{dyn-spectr}}
\end{figure}

\begin{figure}   
\resizebox{\hsize}{!}{\includegraphics[angle=0]{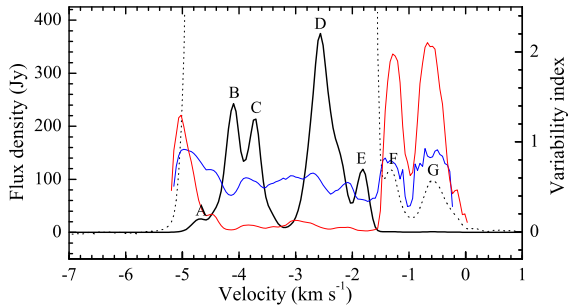}}
\caption{Average 6.7\,GHz maser line profile of Cep\,A (black line) 
with the variability indices $vi_{\rm 1}$ (blue line)
and $vi_{\rm 2}$ (red line) superimposed. The dotted line shows 
the average spectrum magnified by a factor of 50 to display 
flaring features F and G. The mean velocities of labelled features 
are listed in Table \ref{prop-features}.  
\label{average-profile}}
\end{figure}

Changes of the 6.7\,GHz methanol maser spectrum of Cep\,A during 
the monitoring campaign are shown in Fig. \ref{dyn-spectr}. 
There are five features clearly distinguished over the whole observing 
period and two flaring features. They are labelled as A--E 
and F--G, respectively (Figs \ref{dyn-spectr} and \ref{average-profile}). 
The average velocity of these features are listed 
in Table \ref{prop-features}. 

%%%%%%%%%%%%%%%%%%%%%%%%%%%%%%
%%%%%%%%%%%%%%%%%%%%%%%%%%%%%%
%%%%%%%%%%%%%%%%%%%%%%%%%%%%%%
%%%%%%%%%%%%%%%%%%%%%%%%%%%%%%
%%%%%%%%%%%%%%%%%%%%%%%%%%%%%%
\begin{table*}
 \caption{The observed properties of the 6.7\,GHz maser features. 
\label{prop-features}}
\begin{tabular}{c c c c r r r l c}
\hline
Feature  &  Mean        &Variability &Variability     & Flux density  & Period$^a$   & Velocity     &  Component        & Velocity range from \\
         &  velocity    &index 1     &index 2         &range   &              & drift        &  designation from & \citet{torstensson11}  \\
         & (km\,s$^{-1}$)&   $vi_1$   & $vi_2$         &(Jy)    & (day)        & (km s$^{-1}$) & \citet{vlemmings10} & (km s$^{-1}$) \\
\hline
 A       & $-$4.66      & 0.766 & 0.261 &7.0; 82.9 & 84.2$\pm$2.7 & $-$0.10     & 11,12  & $-$4.76; $-$4.32  \\
 B       & $-$4.10      & 0.417 & 0.025 &153; 378 &       & $<$0.03  & 1        & $-$4.32; $-$3.79 \\
 C       & $-$3.72      & 0.559 & 0.065 &90; 325 &        & $-$0.06     & 7,8,9,10& $-$3.96; $-$3.26 \\
 D       & $-$2.60      & 0.611 & 0.062 &152; 590 &  87.5$\pm$7.0 & $+$0.11     & 2,4,5,6 & $-$3.44; $-$1.95   \\
 E       & $-$1.80      & 0.359 & 0.013 &75; 159 &            & $<$0.03  & 3      & $-$1.95; $-$1.51  \\ 
 F       & $-$1.32      & 0.793 & 1.891 &$<$0.3; 6.4 &            & $<$0.03  & -       & \\
 G       & $-$0.55      & 0.898 & 2.031 &$<$0.3; 8.5 &            &$+$0.13     & -     &   \\
\hline
\end{tabular}

\noindent
$^a$ at significance level $p<0.005$ for data from MJD 56057 to 56347.
\end{table*}

The variability index $vi_1$ for spectral channels with flux
densities greater than 1\,Jy ($\cong3\sigma_{\rm rms}$) is plotted
in Fig. \ref{average-profile}. The two persistent features A and D
and flaring features F and G have $vi_1$ greater than 0.61 (Table
\ref{prop-features}).  Feature E has the lowest (0.36) variability index.  
The flaring features F and G have the highest variability index 
$vi_2$ of 1.89 and 2.03, respectively (Fig. \ref{average-profile},
Table \ref{prop-features}). The extreme
blueshifted feature A ($-$4.66\,km\,s$^{-1}$) also exhibits
considerable variability. Note that the variability index reaches a
maximum value of 1.28 near $-$5.05\,km\,s$^{-1}$, i.e. in the blue
wing of feature A. This is due to the outburst activity of feature A
which began after MJD 56083 (Fig. \ref{dyn-spectr}). The lowest
value of $vi_2$ = 0.013 is observed in feature E.

\begin{figure}   
\resizebox{\hsize}{!}{\includegraphics[angle=0]{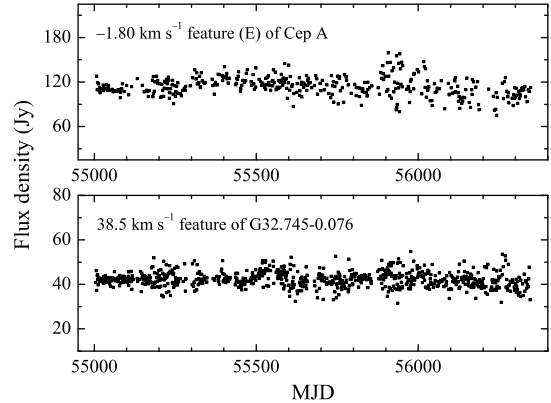}}
\caption{Time series of feature E ($-$1.80\,km\,s$^{-1}$) of Cep~A and 
the strongest feature of G32.745$-$0.076.  
\label{cepa-g32p7-nonvar}}
\end{figure}

The second variability index has the advantage that is sensitive to
flaring and weakly variable features. The first index has rather a
narrow range of values for the features of various levels of
variability. Nevertheless, these two indices are well
correlated. Feature E is one of the permanent features that show
marginal variations in a 3.7\,yr monitoring period. Comparison of its
flux density time series with that of the strongest feature of
the `non-variable' source G32.745$-$0.076 (Fig. \ref{cepa-g32p7-nonvar})
indicates a variation of $\sim$11.5 per cent.  This is only a few
percent higher than the calibration uncertainty. This feature was
previously reported as showing low ($\sim$15 per cent) variability 
on a time-scale of three months \citep{sugiyama08b}.  
   
%%%%%%%%%%%%%%%%%%%%%%%%%%%%%%
%%%%%%%%%%%%%%%%%%%%%%%%%%%%%%

\begin{figure*}   
\resizebox{\hsize}{!}{\includegraphics[angle=0]{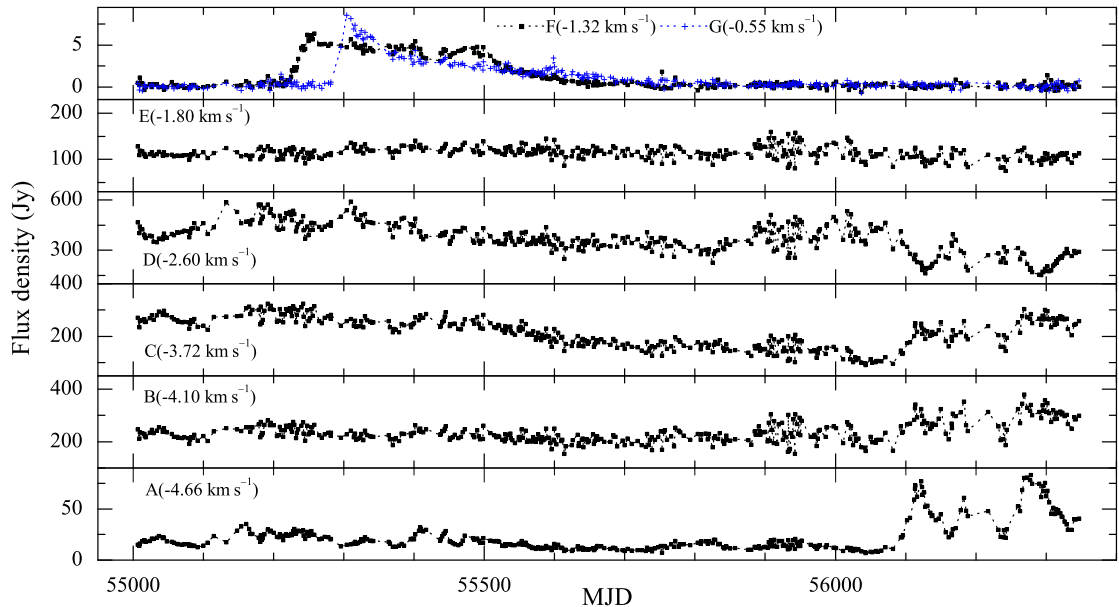}}
\caption{Time series of the flux density of selected features.
\label{feat-var}}
\end{figure*}

Figure \ref{feat-var} shows the flux density of seven features (Table  \ref{prop-features}) 
as a function of time.
There is an indication that the changes in flux density of feature D ($-$2.60\,km\,s$^{-1}$) 
are anticorrelated with the changes in flux density of features A
($-$4.66\,km\,s$^{-1}$), B ($-$4.10\,km\,s$^{-1}$) and C ($-$3.72\,km\,s$^{-1}$) 
for MJD 55007--55115 and 56057--56347. 
This behaviour is clearly visible after MJD 56057
(Fig. \ref{dyn-spectr}), when during $\sim$70\,d, feature D decreased
in flux density by 24 per cent from its initial value, whilst
features C and B increased by 233 and 165 per cent,
respectively. Feature A experienced an increase by a factor of 8.4
(Fig. \ref{feat-var}). Changes in flux density reversed after MJD
56127 when during $\sim$37\,d feature D increased by 86 per cent while
feature A dropped to 250 per cent of the values observed on MJD
56057. No obvious decline in flux density was seen for features B and
C between MJD 56127 and 56164. The synchronized variations of features
A and D continued until the end of the monitoring period. The
flux density of feature D varied inversely to that of features B
and C from MJD 56290 to 56347.  
 
%%%%%%%%%%%%%%%%%%%%%%%%%%%%%%
%%%%%%%%%%%%%%%%%%%%%%%%%%%%%%

\begin{figure}   
\resizebox{\hsize}{!}{\includegraphics[angle=0]{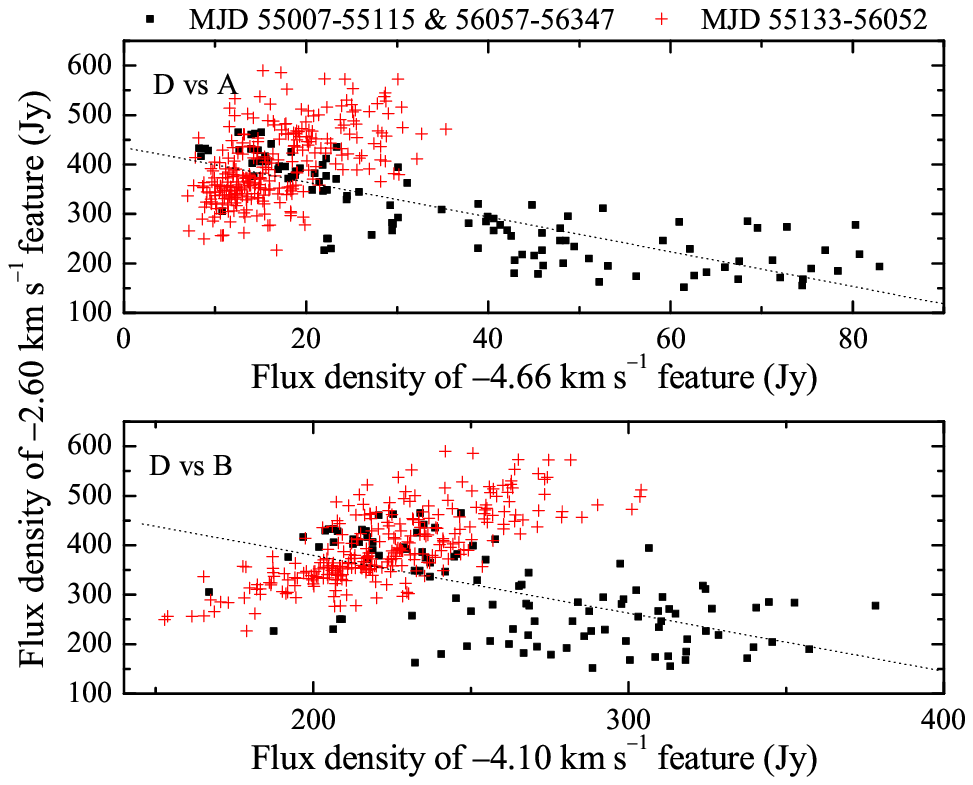}}
\caption{Comparison of flux densities of the strongest redshifted feature D 
with those of blueshifted features A and B. 
Symbols in both panels refer to the time spans given in the legend. 
The dotted lines show the best fits to the data marked by squares.
\label{correlation}}
\end{figure}

Fig. \ref{correlation} shows correlation plots for feature D
versus features A and B. The correlation coefficients were $-$0.82
and $-$0.59, respectively for the data from MJD ranges 55007--55115
and 56057--56347. A positive correlation was seen over a 2.5\,yr
period from MJD 55133 to 56052. Correlation coefficients for the
variation of all the persistent features during the time spans are
listed in Table \ref{corr-coeff}.
We conclude that the synchronized and anticorrelated variations 
occurred for pairs of features A and D, and B and D, i.e. between the blue-
and redshifted emission, for intervals of 110 and 290\,d, which in total is
about 30 per cent of our monitoring period. \citet{sugiyama08b} observed 
similar variations over a 80\,d monitoring period.  

%%%%%%%%%%%%%%%%%%%%%%%%%%%%%%
\begin{table*}
 \caption{Correlation coefficients for different time spans. 
Significant ($p<0.0001$) correlations are shown in bold.
\label{corr-coeff}}
\begin{tabular}{c r r r r c r r r r}
\hline 
      & \multicolumn{4}{c}{MJD 55007--55115 \& 56057--56347} & & \multicolumn{4}{c}{MJD 55133--56052} \\
\cline{2-5} \cline{7-10}
Feature    &  &  &  &  & Feature &   &   &   &  \\
\cline{2-10}
     &  A    & B     & C     & D    &   & A     & B     & C     & D \\
\hline
  B  & {\bf 0.87} &       &       &      &   & {\bf 0.65}  &       &       &   \\
  C  &  0.33 & {\bf 0.47}  &       &      &   & {\bf 0.85}  & {\bf 0.55}  &       &  \\
  D  &{\bf$-$0.82}&{\bf$-$0.59}&$-$0.07&      &   & {\bf 0.55}  & {\bf 0.74}  & {\bf 0.63}  &   \\
  E  &$-$0.06& 0.31  & 0.38  & {\bf 0.56} &   & 0.19  & {\bf 0.77}  & 0.26  & {\bf 0.54} \\
\hline
\end{tabular}
\end{table*}
%%%%%%%%%%%%%%%%%%%%%%%%%%%%%%%

%%%%%%%%%%%%%%%%%%%%%%%%%%%%%%
%%%%%%%%%%%%%%%%%%%%%%%%%%%%%%

After MJD 56057 the emission showed apparent periodic changes (Fig. \ref{feat-var}).  
The Lomb--Scargle periodogram \citep{scargle82} shows  statistically 
significant periods of 84.2 and 87.5\,d for features A and D, respectively 
(Table \ref{prop-features}). No statistically significant periodicities are
found for the other features.

\begin{figure}   
\resizebox{\hsize}{!}{\includegraphics[angle=0]{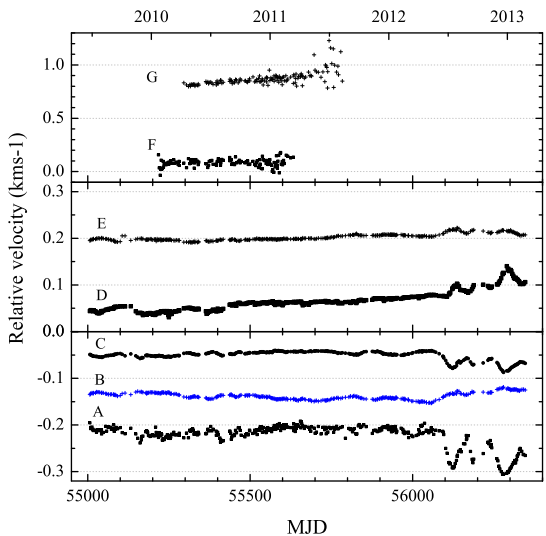}}
\caption{Time series of the relative velocities of peak emission of 
the main features. The labels of features are given in Table \ref{prop-features}.
\label{veldrift}}
\end{figure}

The peak velocities of the main features were measured by parabolic
fitting to the three brightest channels. The values of velocity drifts
are listed in Table \ref{prop-features}. 
The peak velocity of feature A shows a sudden decrease of 0.10\,km\,s$^{-1}$ 
after MJD 56092 (Fig. \ref{veldrift}) with increasing flux density
(Fig. \ref{feat-var}) and the velocity increased with decreasing
flux density. The amplitude of velocity changes during three
consecutive bursts was 0.08\,km\,s$^{-1}$ and the velocity of the peak
during the flux minima decreased by 0.04\,km\,s$^{-1}$ over a period
of 240\,d. These velocity shifts could be caused by variability
in one of a pair of spectrally blended features with almost the same
but slightly different velocities. It is possible that the
lower velocity feature is more stable whilst the higher velocity
feature experiences strong variability (e.g. \citealt{peng89}). This
is plausible explanation because MERLIN data imply that feature A
is composed of two components with velocities which differ by
0.11\,km\,s$^{-1}$ (\citealt{vlemmings10}). Moreover, at some
epochs, this emission was not easily detected in VLBI observations
\citep{torstensson11}.  
Similar amplitude variations were seen for feature C
but with no significant velocity change exceeding 0.05\,km\,s$^{-1}$.
Feature D showed an increase of up to 0.11\,km\,s$^{-1}$ in the peak
velocity, with periodic modulations after MJD 56092
(Fig. \ref{feat-var}), wherein the velocity increased with decreasing
flux density. 
As flaring feature G was relatively weak and well separated
from the rest of the features (Fig. \ref{feat-var}), Gaussian fitting
was used to measure the peak velocity. There was a 0.13\,km\,s$^{-1}$
increase of peak velocity over a time span of 490\,d when its flux
density decreased from 8.7 to $\sim$0.9\,Jy (3$\sigma$
limit). Note that the decline of feature G is correlated with a
lowering of flux density in the rest of the features
(Fig. \ref{feat-var}). We conclude that three persistent features
showed velocity drifts by 2--3 spectral channels. The periodic flux
density variations which are linked to these small velocity drifts argue
against their origin in any systematic radial outflow or infall; they are
more likely to be due to flux density variations in blended components.

%%%%%%%%%%%%%%%%%%%%%%%%%%%%%%

\begin{figure}   
\resizebox{\hsize}{!}{\includegraphics[angle=0]{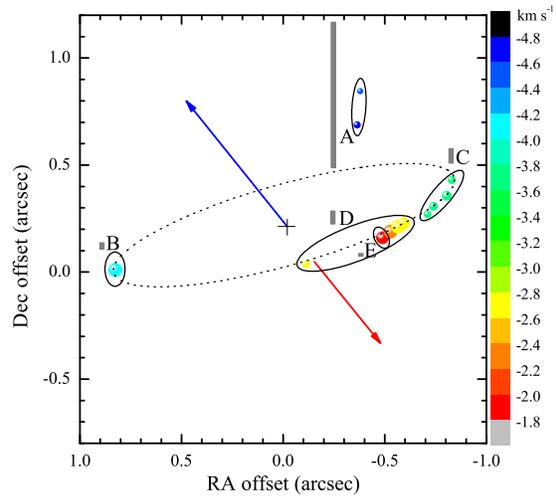}}
\caption{6.7\,GHz maser components in Cep\,A \citep{vlemmings10} with 
variability index information. The groups of maser components marked 
with ellipses and letters correspond to the features identified in 
the single-dish spectra (Table \ref{prop-features}). The length 
of the vertical bars is proportional to the variability index $vi_2$.
The size and colour of symbols correspond to the logarithm of brightness 
and velocity, respectively. The dotted ellipse shows the best fit to 
the maser distribution with the central 7\,mm continuum object 
\citep{curiel06} marked by the cross symbol. The arrows mark 
the blue- and redshifted lobes of the collimated radio 
outflow \citep{curiel06}.
\label{map}}
\end{figure}

Multi-Element Radio-Linked Interferometer (MERLIN) data \citep{vlemmings10} suggest 
that the emission by features identified in the single-dish spectrum comes from spatially
separated components (Fig. \ref{map}). Our features A, C and D consist of up to 
four components (Table \ref{prop-features}) distributed in clumps elongated by 
180$-$570\,mas, which corresponds to projected linear sizes of 130$-$400\,au at 
a distance of 700\,pc \citep{moscadelli09}. VLBI observations 
(\citealt{torstensson11}, their fig. 2 and table 3) suggest that our
feature C is a blend of three components in the velocity range from
$-$3.96 to $-$3.26\,km\,s$^{-1}$, whereas feature D is a blend of
more than three components. The velocity ranges of our individual
features as deduced from the VLBI data are listed in Table \ref{prop-features}. 
We conclude that our observations of features C and D probe 
the variability of multiple spatially different parts of
the ring-like distribution of $\sim$650\,au radius. Their indices
of variability are of intermediate values. Feature A, which has the highest
variability, appeared as a blend of two components in the MERLIN observation 
in 2006 December \citep{vlemmings10} and in the VLBI data from 2004 November 
\citep{torstensson11}. It seems that this emission does not belong to 
the arc-like distribution. Feature B, showing moderate variability, appears 
to correspond to the eastern component with a Gaussian profile 
(\citealt{torstensson11}, their fig. 2). The VLBI data suggest that emission 
from feature E, which has the lowest variability indices, comes from 
a component with a Gaussian profile that is displaced by only $\sim$50\,mas 
from the strongest component of feature D. We conclude that our single-dish 
data provide reliable information for the variability of maser components 
isolated in velocity and unresolved with by the MERLIN beam-size of 40$\times$30\,mas.  

%%%%%%%%%%%%%%%%%%%%%%%%%%%%%%

\subsection{Flares}
Two new emission features F ($-$1.32\,km\,s$^{-1}$) and G ($-$0.55\,km\,s$^{-1}$) 
appeared in the source (Table \ref{prop-features}, Fig. \ref{feat-var}). 
The light curves are rather complex and we obtained the best fits to their rising 
parts using linear function and to their declining parts using exponential
functions ($S_{\nu} \propto {\mathrm e}^{-t/ \tau_{\rm d}}$). The epochs of the
start and the peak of each flare are given in Table
\ref{flar-feat}. The times taken to rise from the $1\sigma_{\rm rms}$
level of $\sim$0.3\,Jy to their peak flux densities, $t_{\rm r}$, were
37 and 24\,d, for features F and G, respectively.  Note that the
rising part of the light curve of feature F was well sampled but for
that of feature G the observations were sparse.  Therefore, the rise
time of feature G is less certain than that of feature F.  Declining
parts of the two light curves were fitted by two different exponential
laws for two intervals $\Delta t_1$ and $\Delta t_2$ (Table \ref{flar-feat}). 
The corresponding decay time-scales $\tau_{\rm d1}$ and $\tau_{\rm d2}$ 
for feature F were 158 and 67\,d while for feature G they were 37 and 251\,d, respectively. 
The flares of emission of features F and G were seen over 513 and 670\,d,
respectively, and the corresponding ratios of the rise time to the
decay time, $R_{\rm rd}$, were 0.08 and 0.04. The time delay between
these flares was $\sim$60\,d.

The rapid rises in the flares were not associated with a rapid rise 
in the total methanol maser luminosity of the source. Flares with similar 
characteristics such as this have not previously been observed to our knowledge
in any methanol source (\citealt*{goedhart04}; \citealt{fujisawa12}). 
A strong OH 1665\,MHz maser flare with a rise time of $\sim$16 months 
was reported from Cep~A \citep{cohen85}. 
Several rapid bursts were also observed in its 22\,GHz water maser 
line with typical rise and decay times of 3--30\,d \citep{mattila85}. 
No MERLIN or VLBI observations are available which show the location 
of the flaring maser components F and G.

%%%%%%%%%%%%%%%%%%%%%%%%%%%%%%
\begin{table*}
 \caption{Parameters of the flaring features. 
\label{flar-feat}}
\begin{tabular}{l c c c c c c c c}
\hline
Feature  &  Epoch of start & Epoch of peak & Amplitude & $t_{\rm r}$ & $\Delta t_1$& $\tau_{\rm d1}$& $\Delta t_2$ &  $\tau_{\rm d2}$ \\
         &  (MJD)       &  (MJD)     & (Jy)      & (d)      & (MJD)     &    (d)       & (MJD)      & (d) \\
\hline
 F       & 55217$\pm$3 & 55254$\pm$3 & 6.3     & 37$\pm$3  & 55254$-$55503$\pm$3&158$\pm$50 &55503$-$55730$\pm$8& 67$\pm$4 \\
 G       & 55281$\pm$5 & 55305$\pm$5 & 8.5     & 24$\pm$7  & 55305$-$55378$\pm$4& 37$\pm$11 &55378$-$55950$\pm$10&251$\pm$15 \\
\hline
\end{tabular}
\end{table*}

%%%%%%%%%%%%%%%%%%%%%%%%%%%%%%
\begin{figure}   
\resizebox{\hsize}{!}{\includegraphics[angle=0]{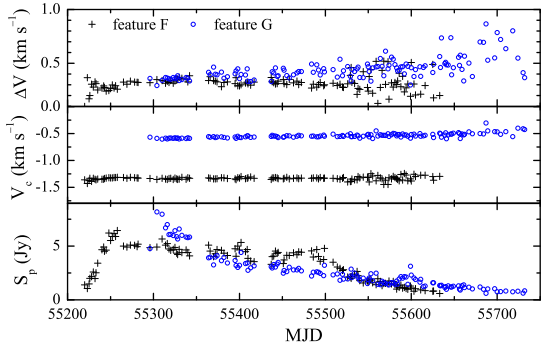}}
\caption{Time series of the peak flux density ($S_{\rm p}$), the velocity of 
the peak ($V_{\rm c}$) and the line-width at half-maximum 
($\Delta V$) of the flaring features F and G.
\label{flare-time}}
\end{figure}
%%%%%%%%%%%%%%%%%%%%%%%%%%%%%%
%%%%%%%%%%%%%%%%%%%%%%%%%%%%%%
\begin{figure}   
\resizebox{\hsize}{!}{\includegraphics[angle=0]{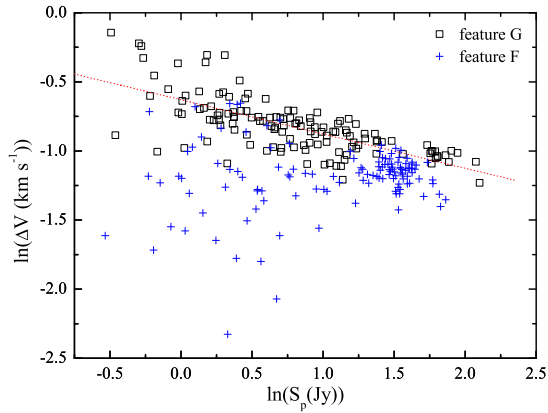}}
\caption{Relationship between the line-width ($\Delta V$) and the flux 
density ($S_{\rm p}$) for the flaring features. 
The dotted line marks the best fit for the feature G.
\label{linewidth-flux}}
\end{figure}
%%%%%%%%%%%%%%%%%%%%%%%%%%%%%%

The profiles of the flaring features are well separated in velocity 
and not blended. We fitted Gaussian curves to the observed
line profiles of features F and G, obtaining the peak flux density, 
$S_{\rm p}$, and the velocity of the peak, $V_{\rm p}$ and
the full width at half-maximum of the line-width, $\Delta V$. 
Fig. \ref{flare-time} shows the variations of these 
three parameters. Feature F did not show significant variations 
in $V_{\rm p}$ and $\Delta V$. 
Note that with decreasing peak flux density the accuracy of fitting 
drops such that the scatter in $\Delta V$ 
gradually increases and individual estimates of less than 
0.2\,km\,s$^{-1}$ and higher than 0.45\,km\,s$^{-1}$ are less reliable.

Feature G exhibited a systematic shift in radial velocity of
0.12\,km\,s$^{-1}$ and an increase in line-width from 0.33 to
0.56\,km\,s$^{-1}$, as the flux density decreased over a time span of 436\,d
during which the parameters could be measured with sufficient
accuracy.  Fig. \ref{linewidth-flux} shows the dependence of the
line-width on the flux density. The least-squares fit to the data of
feature G gives the relationship ln$\Delta V$/ln$S_{\rm p} = -0.25\pm0.02$.  
This can be interpreted as a result of line narrowing during maser 
amplification. The narrowing is a shallower function of intensity than that 
which is predicted by the standard theory of unsaturated masers, i.e. 
ln$\Delta V$/ln$S_{\rm p} = -0.5$ (\citealt{goldreich74}; \citealt{mattila85}).  
The weak dependence of the line-width on the flux density for feature G
and the constant line-width for feature E is poorly consistent with this
theory. One can suggest that either the flares are behaving differently 
from the persistent components and are at least partially saturated,
\emph{and/or} the flares are line-broadened by some other mechanism such
as turbulence (\citealt{vlemmings10}).   

\section{Discussion}
Our densely sampled observations of Cep\,A over the past 3.7\,yr allow
us to study its 6.7\,GHz maser emission behaviour and obtain two main
results.

(i) Synchronized and anticorrelated flux density variations, of moderate
or large amplitude, were seen from the blueshifted features A and B
and the redshifted maser feature D, on time-scales of 110 and 290\,d. 
Periodic variability (84--87\,d) of features A and D was identified 
during the last 290\,d of the monitoring period.

(ii) Two flares, with durations of $\sim$510 and $\sim$670\,d, appeared 
at two different velocities in the redshifted part of the spectrum. 
The time delay between the flares was $\sim$60\,d. None of the persistent 
features A--E in the maser spectrum showed any signs of variability similar 
to the very rapid rise and a slow decay exhibited by features E and G.

The synchronized and anticorrelated variability of methanol maser 
features at the blue- and redshifted velocities in Cep\,A was first 
discovered by \citet{sugiyama08b} during a monitoring period of 81\,d. 
Our observations indicate that this phenomenon of varying intensity 
occurred for the two pairs of features for a total of $\sim$400\,d at 
the beginning and the end of our monitoring period. High angular 
resolution maps of 6.7\,GHz methanol masers  (\citealt{sugiyama08b}; 
\citealt{vlemmings10}; \citealt{torstensson11}) suggest that variability 
is synchronized between different parts of an arched structure. 
In particular, when the flux density of the blueshifted maser components 
to the east and north (features B and A in Fig. \ref{map}) increases, 
the flux density of the redshifted component to the south--west 
(feature D in Fig. \ref{map}) decreases, and vice versa.

The maps of dust continuum, radio continuum and lines (\citealt{curiel06}; 
\citealt{torrelles11}, their fig. 5) imply that in projection the redshifted 
emission is closer to the main heating source \citep{jimenez09} than 
the blueshifted emission. As the methanol maser is probably excited by 
infrared radiation from warm dust (\citealt*{cragg05}), the observed 
variability can be attributed to changes in dust temperature due to 
luminosity variations in the heating source \citep{sugiyama08b}. 
They argued that if the luminosity increases, the dust temperature near 
the source would be too high to sustain the maser emission, whilst still 
being cool enough (100--200\,K) to allow masing further from the source. 
In turn, decreasing luminosity in the heating source would allow masers 
closer to the source but be too cold further away.
This interpretation is poorly consistent with the ring model 
(\citealt{vlemmings10}; \citealt{torstensson11}) where almost the all 
clouds are located at $\sim$650\,au from the central object. 
It appears that at least features A and D (Fig. \ref{map}) lie at
similar distances from the 7\,mm source which is probably the
central exciting object. Changes in dust temperature due to
variability in this object are thus unlikely to cause the
anticorrelated changes in the flux densities of these maser features.

\citet{jimenez07} describe results which suggest a molecular disc
around HW2 with a velocity gradient of $\sim$5\,km\,s$^{-1}$.  If this
is in Keplerian rotation, an observer in a given direction close to
the plane of the disc will see the brightest masers along lines
through the centre of rotation and through the tangential limbs of the
disc (\citealt{elmegreen79}, their fig. 1), giving rise to a
characteristic triple-peaked profile \citep{cesaroni90}. 
This can be explained first by considering isotropic masing in 
the reference frame of the disc; segments separated by 180$^{\circ}$ 
have similar radial velocities with respect to the centre and 
thus maser emission from one side of the disc can be amplified 
further by material on the opposite side. Secondly, from the point 
of view of a distant observer located in a plane close to the disc plane, 
the smallest velocity gradients, and hence, the strongest maser 
amplification in the direction of the observer, 
are located towards the centre and the tangents of the disc.
In this model, the diametrically opposite sections of disc are in
radiative contact with the other. \citet{cesaroni90} have shown that
if the pump rate drops on one side of the disc then the opposite side
will be affected soon after a time equal to the light crossing time.
As a result the blueshifted emission will be anticorrelated with the
redshifted emission or vice versa.  However, this interpretation is
not fully satisfactory because although the VLBI data suggest a ring model,
the dominant maser motion is radial, at 1.3\,km\,s$^{-1}$, not
Keplerian rotation (\citealt{torstensson11}).
The observed velocities of the maser components suggest infall towards
the central object (\citealt{vlemmings10}; \citealt{torstensson11}). 
It may be possible to derive a model for the
synchronized variations of maser features in Cep\,A, including the
effects of changes in one part of the ring on emission in other parts,
if the 3\,D velocity field can be determined from proper motion
measurements.

\subsection{Flaring features}
Flare events in the 6.7\,GHz maser line are very
scarce. \citet{goedhart04} monitored 54 maser sources for 4 yr and
found a flaring feature only in one source, i.e. G351.42+0.64. The
variability pattern of the $-$5.88\,km\,s$^{-1}$ feature in this
source is similar to that of features F and G of Cep\,A; its duration
was $\sim$340\,d, $R_{\rm rd}\approx$ 0.1, it fell exponentially for
two time intervals.  This flare had a maximum flux density of about
230\,Jy, much stronger than the flare maxima of 6.3 and 8.5\,Jy in
Cep\,A. Quasi-periodic flare phenomena of similar patterns but with
much shorter decay time-scales of 5--30\,d and $R_{\rm rd}$ of
0.17--0.20 were reported in methanol masers G33.64$-$0.21 and
G37.55$+$0.20 (\citealt{araya10}; \citealt{fujisawa12}). Other
methanol sources with periodic behaviour are known, such as
G22.357+0.066 with $R_{\rm rd} = 0.34$ \citep{szymczak11} and
G9.62+0.20E with $R_{\rm rd} = 0.37$ \citep*{goedhart03}. Note that,
in periodic and quasi-periodic sources, flaring activity usually shows
correlated variations across several different features in the
spectrum, but the behaviour of flaring features in Cep\,A and
G351.42+0.64 was completely unrelated to the other features. This
indicates that these changes are restricted to a local volume of the
gas.

The emission at velocities of $-$0.53 to $-$0.36\,km\,s$^{-1}$, which
may correspond to feature G, was detected on MJD 53987 in the VLBI
observation but not in the single-dish study \citep{sugiyama08b}. It
appears to coincide with the southern part of the radio continuum jet
of HW2 and with the cluster of maser clumps with velocities of $-$3 to
$-$2\,km\,s$^{-1}$ \citep{sugiyama08b} also seen in the maps taken
with the EVN on MJD 53315 (\citealt{torstensson11}, their fig. 1) and
with MERLIN on MJD 54072 \citep{vlemmings10}. No emission near
$-$0.5\,km\,s$^{-1}$ was detected at these two epochs. This suggests
that feature G undergoes variations of much higher amplitude than
those measured with the 32\,m dish.  Adopting a quiescent value of
0.035\,Jy (the sensitivity limit of the EVN observations;
\citealt{torstensson11}) the flux density of the feature has risen by a
factor of up to 240. The diameter of the cluster is $\sim$40\,au
\citep{torstensson11} for the source distance of 700\,pc
\citep{moscadelli09} and the velocity dispersion of the maser clumps is up
to 2.5\,km\,s$^{-1}$. \citet{torstensson11} argued that the velocity
field of the methanol masers in Cep\,A indicates a modest infall of
maser clumps. The ring structure is probably the interface between the
accretion flow and disc. Direct measurements of infall of methanol
maser clouds towards a high-mass young star were reported by \citet*{goddi11}.

The above discussion may suggest that flaring features F and G arise
in a small region lying in front of the continuum emission of the
radio jet from HW2. The flux density and morphology of this radio
continuum is time variable \citep{curiel06}. In such circumstances,
maser emission from these features can easily be affected by several
factors such as changes in path length, pump rate and background
emission (\citealt{caswell95}; \citealt{cragg05};
\citealt*{vanderwalt09}).

\citet{fujisawa12} proposed a fast ($\sim$1\,d) injection of energy,
released for instance due to magnetic field reconnection, into a small
region, as a plausible mechanism behind the maser flare in source
G33.64$-$0.21. This energy heats the gas and dust, increasing the
infrared flux density which is the main pumping agent of the 6.7\,GHz
transition \citep{cragg05}. Typical magnetic reconnection events are,
however, too short to explain the 24--37\,d rise times of features F
and G. The dust cooling time is only 1.2\,d even in the optically
thick case \citep{vanderwalt09}, which is about two orders of
magnitude shorter than the decay times of the flares we have
observed. Thus, this model cannot be applied to our source.

It is postulated that the velocity field of the 6.7\,GHz masers in
Cep\,A signifies infall motion (\citealt{vlemmings10};
\citealt{torstensson11}). For a flow velocity of 1.3\,km\,s$^{-1}$
\citep{torstensson11}, a maser cloud with a characteristic size of
3\,au and brightness temperature of $3\times10^{12}$\,K \citep{menten92}
could move only 0.4--0.5\,au in 510--670\,d.  This suggests
negligible changes in the cloud environment.

\citet{vanderwalt09} proposed that the declining part of methanol
maser flares in periodic source G9.62+0.20E can be ascribed to a
change in the background photon flux density from a recombining thermal
plasma. They found characteristic decay times from 40 to 400\,d for
electron densities from 10$^6$ to 10$^{5}$\,cm$^{-3}$. This range of
$\tau_{\rm d}$ is fully consistent with the observed values for
features F and G (Table \ref{flar-feat}). The model
of \citet{vanderwalt11} reproduces well the exponential decay of flux
density we have observed.

The location of the flaring maser features is not certain but, as mentioned above, 
it has been suggested  that they lie in the front of radio lobes
with an electron density of about 3--5$\times10^4$ cm$^{-3}$ and
recombination times of 800--1300\,d (\citealt{curiel06}) for a
recombination coefficient of 2.95$\times10^{-13}$\,cm$^{3}$\,s$^{-1}$.  
The radio lobes of HW2 have persisted for a long time but episodes of
ejection separated by 1.85\,yr were observed (\citealt{curiel06}).  
We conclude that the observed characteristics of flaring features F and G
can be explained by variability in the seed photon flux density.  
This scenario could be verified by simultaneous VLBI observations of the maser line
and radio continuum, at epochs guided by single-dish monitoring.

\subsection{Long-term variability}
\begin{figure}   
\resizebox{\hsize}{!}{\includegraphics[angle=0]{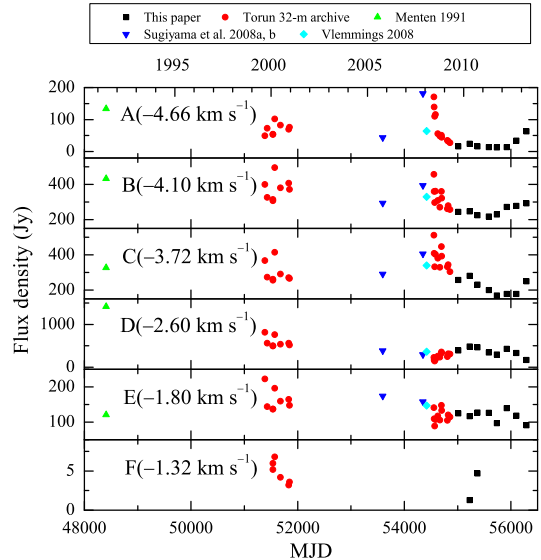}}
\caption{Long-term, $\sim$22\,yr, light curves of main maser features 
labelled as in Fig. \ref{average-profile}. The data before 2009 are 
taken from \citet{menten91}, \citet{sugiyama08a, sugiyama08b}, \citet{vlemmings08} 
and the Torun 32\,m telescope archive. For clarity only some of 
the data presented in this paper (Figs \ref{dyn-spectr} 
and  \ref{feat-var}) are repeated here. Errors in absolute flux
density calibration are within 12--20\, per cent.
\label{longtermvar}}
\end{figure}

The spectra available in the literature (\citealt{menten91};
\citealt{sugiyama08a, sugiyama08b}; \citealt{vlemmings08}), in the
Torun 32\,m telescope archive and seven evenly sampled spectra from
the data set reported in this paper (Figs \ref{dyn-spectr} and
\ref{feat-var}) have been compiled to obtain a long-term light curve.
The discovery spectrum \citep{menten91} was corrected for the
rest frequency of 6668.5192\,MHz \citep*{muller04}. The other archival
spectra have been used as they were published since differences in
rest frequencies have a negligible impact on the velocity scale.  
The light curves of the main features are shown in Fig.
\ref{longtermvar}. Feature D declined by a factor of 5.2 over a period
of 21.6 yr. Its light curve is best fitted by an exponential function
with a decay time-scale of 7.39$\pm$1.39 yr. The feature was clearly
non-Gaussian during the discovery observation (\citealt{menten91}) and
a shoulder at $-$2.3\,km\,s$^{-1}$, corresponding to a second feature,
appeared on MJD 51674.  Feature B shows a significant linear decay of
8.7$\pm$2.0\,Jy\,yr$^{-1}$. A tentative linear decay of
3.1$\pm$1.5\,Jy\,yr$^{-1}$ is observed for feature A. Note that for
this feature the ratio between the maximum and the mean flux density
is 2.80. The peak flux densities of features C and E are fairly steady
over 21.6~yr.  The emission near $-$0.7\,km\,s$^{-1}$, with a peak
flux ranging from 3.2 to 6.8\,Jy, was seen during the MJD range of
51536--51844.  The data imply that the declining part of a flare
event was detected. At those epochs, the precision of radial velocity
measurements was $\pm$0.5\,km\,s$^{-1}$, which makes impossible to
infer whether its peak velocity coincides with that of the
recently detected feature G.

We conclude that the two strongest features, B and D, dropped in flux
density by factors of 1.7 and 5.2 respectively, on a time-scale of
$\sim$22 yr.  The other features do not show significant changes on
this long-term time-scale but they display significant variability on
time-scales of less than 5 yr. The VLBI map \citep{torstensson11}
suggests that the emission of feature D comes from the same cluster of
clouds (clumps {\it c, d, e} in their figs 1 and 2) where from the
emission of stable feature E (clump {\it c}, therein) arises. The
kinematic model of the maser structure in Cep\,A with a radial
velocity of 1.3\,km\,s$^{-1}$, proposed by \citet{torstensson11},
implies that the displacement of maser clouds would be only
$\sim$6\,au after 21.6 yr.  This can be ruled out as the cause of
selective decline of the maser intensity from some clouds in the same
cluster.

\citet{ellingsen07} analysed the 6.7\,GHz methanol maser spectra of 
21 sources and estimated the percentage of spectral features that 
either appear or disappear over a 10\,yr period. He deduced that 
the average lifetime of an individual methanol maser feature is 
approximately 150\,yr. Our comparison of the present and archival 
data imply that feature D showing an exponential decay and features 
B and C with linear decays will disappear after $\sim$50\,yr since the discovery. 
The other persistent features in Cep\,A have probably longer lifetime.

%%%%%%%%%%%%%%%%%%%%%%%%%%%%%%
%%%%%%%%%%%%%%%%%%%%%%%%%%%%%%

\section{Conclusions}
We have studied the 6.7\,GHz methanol maser variability of Cep\,A based 
on 388\,d of monitoring spaced over about 1340\,d. 
We have found  synchronized and anticorrelated flux density variations 
in two blueshifted features A and B, and one redshifted feature D 
over two periods of a total duration of $\sim$400\,d.
This behaviour well constrains causes of variability, excluding shocks 
from an exciting source and supporting the disc model, where the maser 
amplification is switched between radial and tangential modes. 
Although the methanol masers are probably not in a Keplerian disc 
\citep{torstensson11}, they are located in a ring showing either 
outfall or inflow, and changes in excitation conditions could lead 
to maser amplification switching between radial and tangential modes.
Two flares of emission at redshifted velocities (features F and G) 
lasted 510--670\,d. Another flare at a similar
velocity was identified in the archival data. The light curves of the
flares show a rapid rise followed by a slow exponential decline.  
The tentative location of the flaring clouds at the front of the radio jet
from HW2 may suggest that the flares are the results of variability in
the seed flux density. 

These explanation could be tested by simultaneous high resolution 
observations of the maser and radio continuum emission guided by single-dish monitoring.
   
\section*{Acknowledgements}
We would like to thank the referee Simon Ellingsen for constructive comments that improved the manuscript. 
We would also like to thank Anita Richards and Johan van der Walt for their valuable comments and 
Karl Menten and Wouter Vlemmings for kindly sharing their archival spectra in digital form with us. 
The work was supported by the Polish National Science Centre grant 2011/03/B/ST9/00627.

\end{document}